\documentclass[final]{cimento}

\usepackage{epsfig}    
\usepackage[figuresright]{rotating}  

\newcount\plotfigs \newcount\figsonly

\def\CC{\par \vspace*{-2ex} \footnotesize \baselineskip=8pt \begin{verbatim}}

\long\def\startignore #1\stopignore{}   

\def\setlistparams{         
  \topsep=0.7ex                 
  \itemsep=0.7ex                
  \leftmargini=3ex}             
\setlistparams                  

\newcounter{romenumnr}

\newlength{\minipagewidth}

\newsavebox{\boxcontent}
\newcommand{\ovalhead}[1]{
  \unitlength=1cm
  \sbox{\boxcontent}{\mbox{~~{#1}~~}}
  \begin{center}
    \ifdim\wd\boxcontent>6ex 
    \ifdim\wd\boxcontent<8cm 
    \begin{picture}(8,3) \thicklines     
      \put(4.0,0.8){\oval(8,1.6)} 
      \put(0.0,0.7){\parbox{8cm}{
         \begin{center} \usebox{\boxcontent} \end{center}}}
    \end{picture}
    \else \ifdim\wd\boxcontent<12cm 
    \begin{picture}(12,3) \thicklines     
        \put(6.0,0.8){\oval(12,1.6)} 
        \put(0.0,0.7){\parbox{12cm}{
           \begin{center} \usebox{\boxcontent} \end{center}}}
    \end{picture}
    \else
    \begin{picture}(16,3) \thicklines     
        \put(8.0,0.8){\oval(16,1.6)} 
        \put(0.0,0.7){\parbox{16cm}{
           \begin{center} \usebox{\boxcontent} \end{center}}}
    \end{picture}
    \fi \fi \fi
  \end{center}} 

\newcounter{headnr}            
\newcounter{subheadnr}[headnr]
\newcounter{subsubheadnr}[subheadnr]
\def\head #1\par{
  \stepcounter{headnr}                          
  \vspace{2ex} \noindent                        
  {\bf \theheadnr~~~~#1}\\[1ex] \noindent}      
\def\subhead #1\par{  
  \stepcounter{subheadnr}
  \vspace{1.3ex} \noindent
  {\bf \theheadnr.\arabic{subheadnr}~~~#1}\\[0.3ex] \noindent}
\def\subsubhead #1\par{
  \stepcounter{subsubheadnr}
  \vspace{1.0ex} \noindent
  {\bf \theheadnr.\arabic{subheadnr}.\arabic{subsubheadnr}~~~#1}\\ \noindent}

\newlength{\bigcapwidth}
\newlength{\bigcapheight}

\newcounter{stickernr} 
\def\sticker #1\par{ 
  \begin{picture}(100,36.6) 
    \put(0,35){\parbox[t]{90mm}{#1}} 
  \end{picture} 
  \addtocounter{stickernr}{1} 
  \ifnum\value{stickernr}=7 \newpage \setcounter{stickernr}{0} \fi} 

\def\level #1 #2#3#4{$#1 \: ^{#2} \mbox{#3} ^{#4}$}   

\def\la{\mathrel{\hbox{\rlap{\hbox{\lower4pt\hbox{$\sim$}}}\hbox{$<$}}}}
\def\ga{\mathrel{\hbox{\rlap{\hbox{\lower4pt\hbox{$\sim$}}}\hbox{$>$}}}}

\input{epsf}
\def\figspath{}

\def\plotfig#1#2{\epsfxsize=#1
   \ifnum\plotfigs=1 \epsffile{\figspath \figfile.eps}
   \ifnum\figsonly=1 \vspace*{3cm} \par \fi \else \picplace{#2} \fi}
\def\figfile{} \def\viewin {}  \long\def\viewout {}
\ifnum\figsonly=1 \plotfigs=1 \long\def\viewout #1\viewin{\clearpage} \fi
\def\captiontext#1{\ifnum\figsonly=1 \paper
                   \else #1 \fi}  

\def\la{\mathrel{\mathchoice {\vcenter{\offinterlineskip\halign{\hfil
$\displaystyle##$\hfil\cr<\cr\noalign{\vskip1.5pt}\sim\cr}}}
{\vcenter{\offinterlineskip\halign{\hfil$\textstyle##$\hfil\cr<\cr
\noalign{\vskip1.0pt}\sim\cr}}}
{\vcenter{\offinterlineskip\halign{\hfil$\scriptstyle##$\hfil\cr<\cr
\noalign{\vskip0.5pt}\sim\cr}}}
{\vcenter{\offinterlineskip\halign{\hfil$\scriptscriptstyle##$\hfil
\cr<\cr\noalign{\vskip0.5pt}\sim\cr}}}}}	

\def\ga{\mathrel{\mathchoice {\vcenter{\offinterlineskip\halign{\hfil
$\displaystyle##$\hfil\cr>\cr\noalign{\vskip1.5pt}\sim\cr}}}
{\vcenter{\offinterlineskip\halign{\hfil$\textstyle##$\hfil\cr>\cr
\noalign{\vskip1.0pt}\sim\cr}}}
{\vcenter{\offinterlineskip\halign{\hfil$\scriptstyle##$\hfil\cr>\cr
\noalign{\vskip0.5pt}\sim\cr}}}
{\vcenter{\offinterlineskip\halign{\hfil$\scriptscriptstyle##$\hfil
\cr>\cr\noalign{\vskip0.5pt}\sim\cr}}}}}	

\title{Remote sensing of chromospheric magnetic fields\\ \ETC\
via the Hanle and Zeeman effects}

\author{J. Trujillo Bueno\from{ins1}\from{ins2} 
	\atque R. Manso Sainz\from{ins1}\from{ins3}}

\instlist{\inst{ins1} Instituto de Astrof\'{\i}sica de Canarias, 
        E-38205, La Laguna (Tenerife), Spain (jtb@iac.es)
        \inst{ins2} Consejo Superior de Investigaciones Cient\'\i ficas, Spain
	\inst{ins3} Dipartimento di Astronomia e Scienza dello Spazio,
	Universit\`a degli Studi di Firenze, $\,\,$ 
	I-50125, Florence, Italy (rsainz@iac.es)}

\PACSes{\PACSit{96.60.-j}{Solar physics}
	\PACSit{95.30.Gv}{Radiation mechanisms; polarization}
	\PACSit{95.30.Jx}{Radiative transfer; scattering}
	\PACSit{32.80.Bx}{Level crossing and optical pumping}
	   \PACSit{01.30.Cc}{Invited review in conference proceedings}}

\begin{document}

\maketitle

\begin{abstract}

The only way to obtain reliable empirical information 
on the intensity and topology of the
weak magnetic fields of the ``quiet'' solar chromosphere
is via the measurement and rigorous physical interpretation of
polarization signals in chromospheric spectral lines.
The observed Stokes profiles reported here are due
to the Hanle and Zeeman effects operating in a weakly magnetized
plasma that is in a state far from local thermodynamic equilibrium.
The physical origin of their ``enigmatic'' linear 
polarization $Q$ and $U$ components
is the existence of atomic polarization in their metastable lower-levels,
which permits the action of a dichroism mechanism that has nothing to do
with the transverse Zeeman effect. 
It is also pointed out that the population imbalances and coherences
among the Zeeman sublevels of such long-lived atomic levels can 
survive in the presence of horizontal magnetic fields
having intensities in the gauss range, and produce significant polarization 
signals. Finally, it is shown how the most recent developments in the
observation and theoretical modelling of weak
polarization signals are facilitating fundamental new advances
in our ability to investigate the magnetism of the outer
solar atmosphere via spectropolarimetry.

\end{abstract}

\section{Introduction}

The physical processes that
underlie solar magnetic activity are of fundamental importance to astrophysics
as well as in controlling the heliosphere including near-earth space weather. 
However, with the possible exception of the solar photosphere ---the thin 
surface layer where almost all of the radiative energy flux is emitted---, 
our empirical
knowledge concerning the magnetism of the outer solar atmosphere 
(chromosphere, transition region, corona) is still very primitive.
This is very regrettable because
many of the physical challenges of solar and stellar 
physics arise precisely from magnetic processes
taking place in such outer layers. 

In particular, the ``quiet'' solar chromosphere
is a crucial region whose magnetism we need to understand 
for unlocking new discoveries.
It is in this highly inhomogeneous and dynamic
region of low density plasma overlying the thin solar photosphere where 
the magnetic field becomes the globally dominating factor. If we
aim at understanding the complex and time-dependent structure
of the outer solar atmosphere 
we must first decipher how is the intensity and
topology of the magnetic fields of the solar chromosphere.

According to the ``standard picture'' of chromospheric 
magnetism described in the recently-published
Encyclopedia of Astronomy and Astrophysics
there is ``a layer of magnetic field which
is directed parallel to the solar surface and located
in the low chromosphere, overlying a field-free region
of the solar photosphere''. This so-called {\it magnetic canopy}
``has a field strength of the order of 100 gauss and covers a
large fraction of the solar surface'' \cite{25}.

This picture of chromospheric magnetism
seems to be in the minds of most solar physicists
since the beginning of the 1980s, when R.G. Giovanelli and H.P. Jones
interpreted solar magnetograms in chromospheric lines (like the IR triplet
of ionized calcium or the Mg I $b_2$ line) taken in network unipolar regions
near the solar limb, as well as in sunspots and related active regions.
Such chromospheric magnetograms seem to show a polarity inversion
and are considerably more diffused in appearence than photospheric 
magnetograms,
which is interpreted as the result of the expansion
of the magnetic field lines with height in the solar atmosphere
\cite{2, 23, 10, 11, 13}.

The magnetic canopy model was later reinforced in the 1990s via magnetohydrostatic
extrapolations of photospheric magnetic flux tube models 
\cite{24}.
However, it was found that only magnetic field extrapolations that allow
for substantial differences between the temperatures 
of the atmospheres within and outside the 
assumed magnetic flux tubes are capable of producing a 
low-lying canopy field. 
If the internal and external atmospheres are assumed to be 
similar the canopy extrapolated field forms in the upper chromosphere and the corona.
It was argued that the assumption of much lower temperatures in the
external atmosphere fits nicely with the observational finding 
of strong CO absorption lines near the extreme solar limb \cite{1}. 

Magnetograms and extrapolations 
thus led to the idea that the ``quiet'' solar chromosphere 
is pervaded by magnetic canopies with predominantly horizontal fields
overlying ``field-free'' regions whose temperatures 
remain relatively cool up to the canopy bases. As a matter of fact,
some researchers investigated
the impact of ``the magnetic canopy'' on the frequencies of solar $p$- and $f$-modes
(see {\em e.g.}, \cite{7}), while others found of interest to consider
its influence on the linear polarization of some resonance lines 
\cite{8}.
It is however very important to emphasize that,
as pressed with great force by a working group on chromospheric 
fields (see \cite{12}), chromospheric magnetograms 
have never ``detected'' magnetic canopies in the truly
quiet Sun where the network is fragmentary and
photospheric magnetograms show the well-known
``salt and pepper'' patterns of mixed polarity. 
In fact, the Ca {\sc ii} IR triplet and other chromospheric lines are relatively broad, 
which implies that the magnetic fields of the ``quiet'' chromospheric
regions are difficult to diagnose via the only consideration
of the longitudinal Zeeman effect on which magnetograms are based on.
Obviously, the above-mentioned chromospheric magnetograms
(of network and active regions)
and magnetohydrostatic extrapolations 
(of photospheric magnetic flux tube models) are not suitable
for drawing conclusions on the magnetism of the
most quiet regions of the solar chromosphere. 

Over the last few years, observational investigations of scattering
polarization on the Sun have pointed out the existence of ``enigmatic''
linear polarization signals in several spectral lines (observed in the
``quiet'' solar chromosphere close to the limb
as well as in solar filaments), which
cannot be understood in terms of the classical theory of scattering polarization
\cite{30, 19, 31, 37, 39}.
In particular, the ``enigmatic'' features of the
linearly-polarized solar-limb spectrum have motivated some novel theoretical
investigations of scattering polarization in spectral lines
\cite{35, 17, 18, 33, 34, 20, 38, 39}.
Such investigations have been carried out within 
the framework of polarization transfer theories that
allowed us to formulate scattering polarization problems
taking into account a physical ingredient that had been previously
neglected: ground-level atomic polarization
({\em i.e.}, the existence of population imbalances and/or coherences
among the Zeeman sublevels of the lower-level of the spectral line
under considerartion).

Of particular interest in this respect is the letter published in Nature by
Landi Degl'Innocenti with the title ``Evidence against
turbulent and canopy-like magnetic fields in the solar chromosphere'' 
\cite{17}.
He concludes that the explanation in terms of ground-level atomic polarization of 
the ``enigmatic'' linear polarization peaks of the sodium D-lines
observed by Stenflo and Keller in quiet regions close to the
solar limb \cite{30}, implies that 
the magnetic field of the ``quiet'' solar chromosphere has to be
either isotropically distributed but extremely low (with $B {\la} 10$ milligauss)
or, alternatively, practically vertically orientated. 
More recently, the opinion that magnetic fields of milligauss or weaker 
strength cannot exist in the highly conductive solar atmospheric plasma 
has led Stenflo {\em et al.} 
to the conclusion that the magnetic field in the most quiet regions 
of the solar chromosphere has then to be preferentially vertical \cite{32}.

The only way to obtain reliable empirical information 
on the intensity and topology of the
weak magnetic fields of the ``quiet'' solar chromosphere
is via the measurement and rigorous physical interpretation of
weak polarization signals in chromospheric spectral lines.
The aim of this keynote article is to show in some detail how the most recent advances
in the observation and physical interpretation of weak
polarization signals in terms of the Hanle and Zeeman effects
is giving us decisive new clues about the topology and intensity
of the magnetic fields of the ``quiet'' solar chromosphere.

\section{The Zeeman and Hanle effects}

In order to understand why the observed polarization signals 
reported in Section 3 are weak, 
first we need to advance something concerning their physical origin.
The {\it circular} polarization signals
are mainly due to the
longitudinal Zeeman effect. As is well known, Zeeman-induced circular
polarization signals are sensitive to the net magnetic flux density
over the spatio-temporal resolution element of the observations. 
Although it is true that a complex magnetic field topology within the line formation region
may conspire to make the observed circular polarization signals weak,
we have some good reasons to believe that the
Stokes $V$ signals are weak
mainly because the magnetic fields of the ``non-magnetic'' 
solar chromosphere are intrinsically weak ({\em i.e.}, below 100 gauss).

The physical origin of the observed {\it linear} polarization signals 
is completely different and 
has nothing to do with the transverse Zeeman effect.
The observed Stokes $Q$ and $U$ signals are due to {\it atomic polarization}, 
{\em i.e.}, to the existence
of population imbalances and quantum interferences (or coherences) 
among the sublevels pertaining to the upper and/or lower 
atomic levels involved in the line transition under consideration.
This atomic polarization 
is the result of a transfer process of ``order'' from the radiation
field to the atomic system (see \cite{34}). The most obvious
manifestation of ``order'' in the solar radiation field
is its degree of anisotropy arising from its centre-to-limb variation.
In fact, the main source of atomic polarization is the anisotropic
illumination of the atoms of the solar atmospheric plasma,
which produces a {\em selective} radiative pumping.
This pumping is ``selective'', in the sense
that it produces population imbalances among the Zeeman sublevels of each
atomic level. This implies sources and sinks of linear (and even circular) polarization
at each point within the medium. These locally generated polarization signals
are then modified via transfer processes in the stellar plasma. The emergent
polarization signals are weak because the degree of anisotropy
of the solar radiation field is weak (which leads to population imbalances
and coherences that are small compared with the overall population of the atomic level
under consideration), but also because we have collisions and magnetic fields
which tend to modify the atomic polarization.

The Hanle effect is the modification of the atomic polarization (and of the
ensuing {\it linear} polarization profiles $Q(\lambda)$ and $U(\lambda)$) 
due to the action of a weak magnetic field
(see the review \cite{34}). 
As the Zeeman sublevels of degenerate 
atomic levels are split by the magnetic field, the degeneracy is lifted and, 
as long as the sublevels still overlap, 
the coherences 
(and, in general, also the population imbalances 
among the sublevels) are modified. 
Therefore, the Hanle effect is sensitive 
to magnetic fields such that the corresponding Zeeman splitting is comparable 
to the inverse lifetime (or natural width) of the lower or the upper 
atomic levels of the line transition under consideration.
On the contrary, the Zeeman effect is most sensitive in {\em circular}
polarization (quantified by the Stokes $V$ parameter), with
a magnitude that scales with the ratio between the Zeeman splitting
and the width of the spectral line (which is very much larger than
the natural width of the atomic levels). 

The basic approximate formula to estimate
the {\it maximum} magnetic field intensity $B$ (measured in gauss)
to which the Hanle effect can be sensitive is
\begin{equation}\label{1}
10^6\,\,{B}\,\,g_{J}\,\,\approx\,\,1/t_{\rm life}\,\,,
\end{equation}
where $g_{J}$ and $t_{\rm life}$ are, respectively, the Land\'e factor 
and the lifetime in seconds of the atomic
level under consideration (which can be either the upper or the lower level 
of the chosen spectral line transition). 
This formula shows that the measurement
and physical interpretation of weak polarization signals in suitably chosen spectral
lines may allow us to diagnose magnetic fields having intensities between
$10^{-3}$ and 100 gauss approximately, {\em i.e.}, in a parameter domain that is very hard
to study via the Zeeman effect alone. 

While the Hanle effect modifies the atomic polarization,
elastic collisions always produce atomic-level {\it depolarization}. The depolarization
is complete only if $D\,t_{\rm life}\,{\rightarrow}\,{\infty}$,
where $D$ (given in ${\rm s}^{-1}$) is the depolarizing rate of the
given atomic level and $t_{\rm life}$ its lifetime.
Therefore, at first sight, one would be tempted to
conclude that {\it ground} and {\it metastable} levels 
are more vulnerable to elastic collisions than atomic levels of shorter lifetimes.
This is however only true if $D$ is assumed to be of the same order-of-magnitude for
both, the long-lived and short-lived atomic levels under consideration. 
Unfortunately, our current knowledge on depolarizing rates due to elastic
collisions is very poor, but we may hope to use the Sun itself
as an atomic physics laboratory for improving the situation.

\section{Observations of weak polarization signals in chromospheric lines}

Stenflo and Keller \cite{30} have adopted the term 
``the second solar spectrum''
to refer to the linearly polarized solar limb spectrum
which can be observed with spectropolarimeters that allow the detection
of very low amplitude polarization signals (with $Q/I$
of the order of $10^{-3}$ or smaller). Such observations 
with the polarimeter ZIMPOL (see also the atlas \cite{9})
have been confirmed (and extended to the full Stokes vector) by
Dittmann {\it et al.} \cite{5}, Mart\'\i nez Pillet {\it et al.} \cite{22}
and Trujillo Bueno {\it et al.} \cite{37} using the Canary Islands 
telescopes. One of these telescopes is TH\'EMIS, which has 
allowed us to carry out observations of the full Stokes vector in several spectral
lines simultaneously \cite{37}. 
Given the increasing interest of this research field, 
TH\'EMIS is being used also by many other colleagues 
(see, {\em e.g.}, Bommier's report in this volume).
Thanks to a modified version of their polarimeter, 
Stenflo {\em et al.} have also
started to investigate the four Stokes parameters of optical spectral lines
in solar regions near the limb with varying degrees of magnetic activity
\cite{32}.
It is also of interest to point out the enormous diagnostic potential
offered by the near-UV spectral region where the degree of anisotropy
of the solar radiation field is relatively high. Fortunately, there is at least
one solar polarimeter that has been developed recently thinking seriously in
the scientific interest of this near-UV region: ZIMPOL-UV.  

In the remaining part of this section we show some particularly interesting 
examples of our own spectropolarimetric observations
in optical and near-IR chromospheric lines, which have been obtained using different
polarimeters attached to the Tenerife solar telescopes (VTT, GCT and TH\'EMIS).
As we shall see below, 
the physical interpretation of these observations in terms
of the quantum theory of polarization highlights the key role played
by some subtle physical mechanisms in producing the emergent
polarization.

Figure 1 shows an example of our VTT+TIP observations using
the He {\sc i} 10830 \AA$\,$ multiplet \cite{39}.
TIP is the Tenerife Infrared Polarimeter, which is based on ferroelectric-liquid-crystals \cite{21}.
The figure shows the case of a solar filament that was located {\it exactly} at the
very center of the solar disk during the observing day.
The open circles indicate the spectropolarimetric observation, while the solid line
shows the theoretical modelling based on the density matrix
polarization transfer theory (see Section 5). 
Like prominences, solar filaments are
magnetized plasma ribbons embedded in the $10^6$ K solar corona, and
confined by the action of highly inclined magnetic fields (with respect
to the stellar radius) and having intensities in the gauss range. 
(The only difference is that prominences are observed off-the-limb,
{\em i.e.}, against the dark background of the sky, while filaments
are observed against the bright background of the solar disk.
Therefore, we see emission lines in prominences,
but absorption lines in filaments.)

\viewin \def\figfile{jtbueno_1}
 \begin{figure}[t]
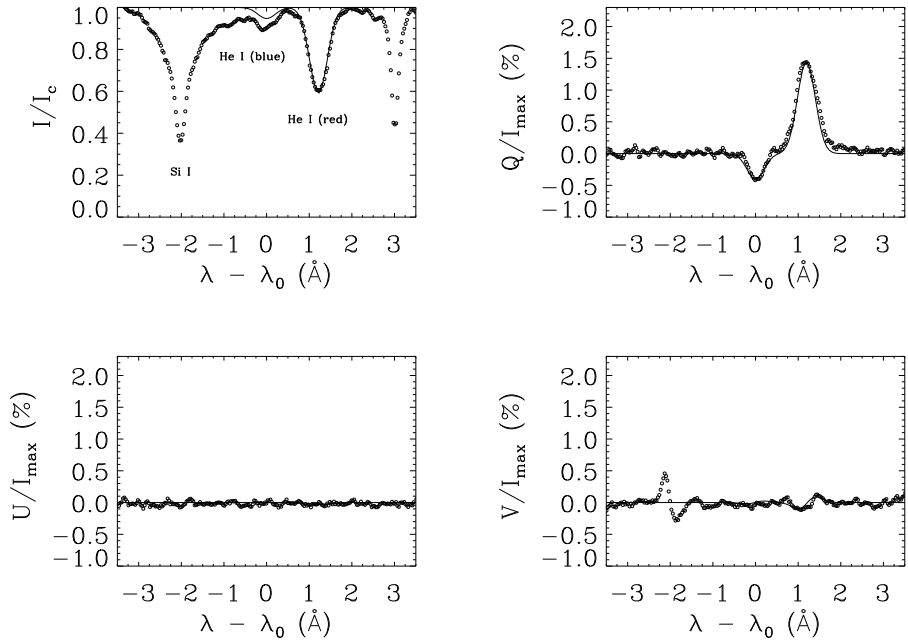

 \plotfig{130mm}{130mm}
   \caption[]{\label{fig:one} \captiontext{
He {\sc i} 10830 \AA$\,$ spectropolarimetric observation of a solar filament located
at the solar disk center (open circles) versus theoretical modelling
(solid line). The fit to the observations has been achieved assuming
a magnetic field vector of 20 gauss inclined
by 105$^\circ$ with respect to the radial direction and 
that the observed filament region was located at a height of $40''$
above the solar photosphere.
The positive reference direction for Stokes $Q$ is parallel
to the projection of the magnetic field vector on the solar disk.
It turns out that this direction made an angle of
about 10$^\circ$ in the clockwise direction with respect to the axis
of the solar filament. The Stokes parameters are normalized to the
maximum line-core depression (from the continuum level) of the
Stokes $I$ profile of the ``red'' absorption line. This observation
has been obtained with the TIP polarimeter attached to the
Vacuum Tower Telescope (VTT). From \cite{39}.}}
\end{figure} \viewout

The observational results of Fig. 1 are very interesting.
First of all, we have sizable Stokes $Q$ signals in both the
``blue'' and ``red'' components of the He {\sc i} 10830 \AA$\,$ multiplet.
This demonstrates that the Hanle effect 
can give rise to significant linear polarization
even at the very center of the solar disk where we meet the case of
forward-scattering (see \cite{34}).
Moreover, the fact itself that
the ``blue'' component is linearly polarized is particularly interesting
because it is the result of $J_l=1\,{\rightarrow}\,J_u=0\,{\rightarrow}\,J_l=1$
scattering processes (with $J_l$ and $J_u$ the total angular momentum
of the lower and upper levels, respectively). 
According to  
scattering polarization transfer theories neglecting the role of lower-level
atomic polarization,
such a line transition should be
intrinsically unpolarizable because the upper level, having $J_u=0$,
cannot carry any atomic polarization. We will see below
that the physical origin of this ``enigmatic'' linear polarization signal
is the existence of a sizable amount of atomic polarization in the
lower level, whose $J_l=1$. 

Another interesting feature of our
solar filament observation is that the ``blue'' and ``red''
lines of the He {\sc i} 10830 \AA$\,$ multiplet show up with amplitudes
of opposite sign, which cannot be modeled via the assumption of
independent two-level atomic models for the three line components
of the helium multiplet. At least
a two-term atom taking into account the fine structure of the upper
$2^3P_{2,1,0}$ term is needed in order to be able to obtain
qualitative agreement with the observed linear polarization amplitudes.

\begin{figure}
  \centerline{\epsfig{figure=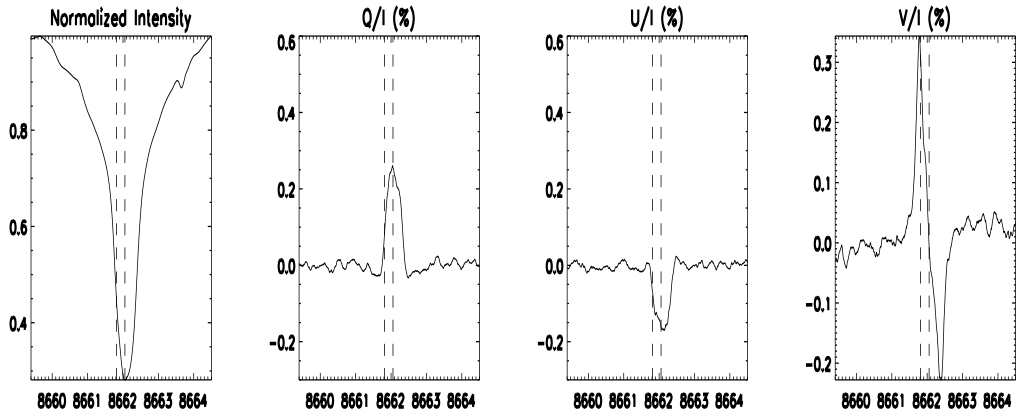, width=6cm, height=14.25cm, angle=90}}
  \caption{\label{fig:two} 
	The full Stokes vector of the Ca {\sc ii} 8662 \AA$\,$ line
observed on the solar disk at about 5'' from the limb during the
equinox period of September 2000. The positive reference direction for Stokes $Q$
is along the line perpendicular to the radial direction through the observed point. 
The vertical dashed-line to the {\it rhs}
of each panel indicates the central wavelength of the 8662 \AA$\,$ line,
while the {\it lhs} dashed-line gives the position of a nearby photospheric iron line. 
This spectropolarimetric observation with the Tenerife Gregory Coud\'e Telescope (GCT)
is the result of a
collaboration between Dittmann, Semel and Trujillo Bueno.}
\end{figure}

Figure 2 shows
the full Stokes vector of the Ca {\sc ii} 8662 \AA$\,$ line observed
on the disk at 5'' from the solar limb. This observation is the result
of a collaboration between Dittmann, Semel and Trujillo Bueno. 
They have used Semel's stellar polarimeter attached to the Tenerife 
Gregory Coud\'e Telescope
and carried out during September 2000
spectropolarimetric observations of the Ca {\sc ii} IR triplet
in regions near the limb with varying degrees of magnetic activity.
The Ca {\sc ii} 8662 \AA$\,$ line is of particular interest because
its upper level, having $J_u=1/2$, cannot harbour any atomic alignment.
This has led to consider the reported 
detection of a significant Stokes $Q/I$ amplitude in this spectral line 
as ``enigmatic'',
because of the belief that the 
polarization effects come only from the population imbalances and coherences 
in the {\it excited} states of the scattering process \cite{31}. 
The full Stokes vector observation of Fig. 2
shows the existence of sizable linear polarization signals 
in the Ca {\sc ii} 8662 \AA$\,$ line, both in $Q/I$ and $U/I$. 

Finally, Fig. 3 shows an example of TH\'EMIS observations
of the second solar spectrum (see \cite{37}).
It shows the full Stokes vector of the oxygen IR triplet at 777 nm, 
as observed on-the-disk
at 4'' from the North solar limb. It is of great scientific interest to point out that
the two lines at 7772 \AA $\,$ and 7774 \AA $\,$ have {\it positive}
$Q/I$ fractional linear polarization amplitudes, while the
7776 \AA $\,$ line shows {\it negative} polarization ({\em i.e.}, along the
solar radius through the observed point!) 
all over its full spectral range. 
A forthcoming publication
will show in detail that this is due to the existence
of atomic polarization in the {\it metastable} lower-level of the oxygen 
triplet\footnote{This oxygen triplet is a beautiful example of the case
of three lines with a common {\it lower} level. The case
of three lines with a common {\it upper} level ({\em e.g}. the Mg {\sc i} $b$ lines),
has been investigated by Trujillo Bueno, concluding
that the physical origin of the ``enigmatic'' $Q/I$ amplitudes observed
by Stenflo {\em et al.} \cite{26, 31} is, again, the atomic polarization of
their metastable lower levels \cite{33, 34}.}.
Finally, note that in these oxygen lines we find
significant circular polarization signals, which can only be produced by
magnetic fields substantially larger than 0.01 gauss, as is the case also with the
Stokes $V$ profiles of the Ca {\sc ii} 8662 \AA$\,$ line shown in Fig. 2.

\viewin \def\figfile{jtbueno_3}
 \begin{figure}[t]
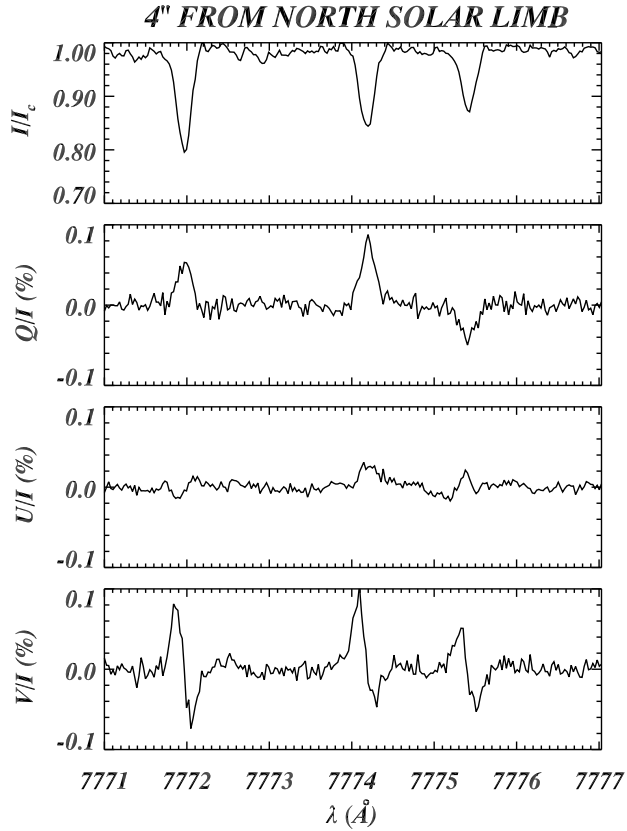

 \plotfig{110mm}{110mm}
 \caption[]{\label{fig:three} \captiontext{
The fractional polarization of the 
oxygen IR triplet at 777 nm observed with TH\'EMIS
on the solar disk at about 4'' from the North solar limb. 
From \cite{37}.}}
\end{figure} \viewout

\section{The physical origin of the enigmatic polarization signals:
atomic polarization of metastable levels and dichroism}

The ``enigmatic'' signals of the ``second solar spectrum''
are detected in spectral lines whose lower level is the ground
state or a metastable level. These levels have a lifetime
$t^{l}_{life}{\approx}1/B_{lu}\bar{J}^0_0$, which is much larger
than the upper level lifetime $t^{u}_{life}{\approx}1/A_{ul}$
(with $B_{lu}$ and $A_{ul}$ are the Einstein coefficients 
for absorption and stimulated emission, respectively, 
and $\bar{J}^0_0$ is the line integrated mean intensity
of the radiation field).
Therefore, the atomic polarization of
such {\em long-lived} lower-levels is very sensitive to depolarizing mechanisms. 
However, it is very important to emphasize that only collisions can
depolarize completely a given atomic level. Except
for a few very particular cases (see \cite{40}), the depolarization of the atomic levels due
to magnetic fields (the Hanle effect) is never complete. 
For instance, elastic collisions and a microturbulent and isotropic magnetic field
modify the degree of population imbalance of the upper level
of a two-level atom (with $J_l=0$ and $J_u=1$) as dictated
by the following approximate expression (cf. \cite{36}):
\begin{equation}
\sigma^2_0\,=\,{{\rho^2_0}\over{\rho^0_0}}\,\,{\approx}\,\,{{\cal H}\over{1+\delta}}\,\,{\cal A},
\end{equation}
where ${\cal A}=\bar{J}^2_0/\bar{J}^0_0$ is the anisotropy 
factor\footnote{Its possible values are such that $-{{1}\over{2}}\,\la\,
{\sqrt{2}{\cal A}}\,\la\,1$. (Note that there is a typing error in 
Eq. (11) of \cite{34}, since the 
inequalities given there are correct for $2{\cal A}$, {\it not}
for ${\cal A}$.)} 
of the pumping radiation field,
$\delta$ is the collisional depolarizing rate in units of the Einstein
$A_{ul}$ coefficient, and $\cal H$ is the Hanle depolarization factor
which varies between 1 (for the zero magnetic field case) and $1/5$
(for a Zeeman splitting very much larger than the natural width
of the upper level). We point out that $\sqrt{3}{\rho^0_0}$
is the {\it overall} population of the upper level, while
its atomic alignment (or degree of population imbalance) is quantified
by ${\rho^2_0}=[N_1\,-2N_0\,+N_{-1}]/{\sqrt{6}}$ (where $N_i$ are the
individual populations of the three Zeeman sublevels of the upper level). 

A key question is the 
following\footnote{In this article, the atomic polarization 
of a given atomic level is quantified
by means of the spherical tensor components of its atomic density
matrix and the quantization-axis (the $z$-axis) is 
taken along the stellar radius (see \cite{34}).}: 
to which extent can the atomic 
polarization of long-lived atomic levels survive the partial Hanle-effect
destruction produced by highly inclined magnetic fields having intensities
in the range of gauss ?. The answer to this question is
of greatest importance for a correct diagnostics of the magnetic
fields of the outer solar atmosphere (chromosphere, transition region
and corona). This is because, as clarified below, the physical origin of the above-mentioned
``enigmatic'' polarization signals is the existence of a significant
amount of population imbalances and coherences in the {\it metastable} lower-levels
of their respective spectral lines. 

Interestingly, the upper level of some of the
``enigmatic'' spectral lines cannot carry any atomic alignment (because
it has $J_u=0$ or $J_u=1/2$, as is the case with the ``blue'' line of the
He {\sc i} 10830 \AA $\,$ multiplet of Fig. 1 or with the Ca {\sc ii} 8662 \AA $\,$ 
line of Fig. 2, respectively). For this type of lines there is no contribution of
upper-level atomic polarization to the $Q$ and $U$ components
of the {\it emission} vector ({\em i.e.}, to $\epsilon_Q$ and $\epsilon_U$),
simply because such upper levels cannot carry any atomic alignment.
Moreover, the contributions to $\epsilon_Q$ and $\epsilon_U$
arising from the Zeeman splitting of the lower level ({\em i.e.}, due to the
transverse Zeeman effect) are 
negligible for the ``weak'' magnetic fields of
prominences, filaments and of the ``quiet'' solar chromosphere and corona.
This type of lines (with $J_u=0$ or $J_u=1/2$)
may thus be called ``null'' lines, because the spontaneously emitted
radiation that follows the anisotropic radiative excitation is vitually {\it unpolarized}.

However, $\eta_Q$ and $\eta_U$ can have sizable values, if a significant amount of
lower-level atomic polarization is present. In principle, this is possible
for the aforementioned helium and calcium lines because their lower levels 
can be polarized (because they have $J_l$=1
and $J_l=3/2$, respectively). When a sizable amount of atomic polarization
is present in such lower levels, as it happens
in the outer solar atmosphere, then the role of the emissivity in 
Stokes $Q$ and $U$ comes exclusively from the terms $-\eta_Q\,I$ and $-\eta_U\,I$
that arise in their respective radiative transfer 
equations. If the Stokes-$I$ intensity along the line of sight is important enough
(as it occurs for the on-the-disk observations of Figs. 1,2 and 3),
then we can have an important contribution of the absorption process
itself to the emergent linear polarization. We call this mechanism 
{\it dichroism}
in a weakly magnetized medium which, we would like to stress, 
has nothing to do with the
transverse Zeeman effect (see \cite{35}).
This {\it dichroism} mechanism, which requires the presence of a sizable amount
of lower-level polarization, plays a crucial role in producing the observed
``enigmatic'' linear polarization signals in a variety of chromospheric
lines \cite{33, 34, 20, 38, 39}.

The conclusion that some of the ``enigmatic'' linear polarization 
signals are due to {\it dichroism} demonstrates that a sizable amount
of atomic polarization is present in the lower levels of such spectral lines.
As mentioned above, such lower-levels are metastable 
({\em i.e.}, they are long-lived atomic levels). According to the basic Hanle-effect Eq. (1)
their atomic polarization is vulnerable to magnetic fields of very low
intensity ({\em i.e.}, to fields ${\rm B}\,\ga\,10^{-3}$ gauss !). 
This magnetic depolarization takes place
for sufficiently inclined fields with respect to the radial direction
of the star ({\em i.e.}, for $\theta_B{\ga}$10$^\circ$). Unfortunately,
the particular conclusion of Landi Degl'Innocenti that the atomic
polarization of the hyperfine components of the ground level of sodium does not
survive sufficiently in the presence of turbulent or canopy-like horizontal fields
stronger than about 10 milligauss \cite{17} has led to unjustifiable reinforcements
of the belief that the atomic polarization of {\it any} long-lived
atomic level has to be insignificant in the presence of highly inclined
solar magnetic fields having intensities in gauss range 
\cite{27, 28, 29}. If this belief
were correct in general, then it would be justified to conclude that
the magnetic field throughout much of the ``quiet'' solar chromosphere
has to be either extremely low (with ${\rm B}\,{\la}\,0.01$ gauss) or,
alternatively, oriented fairly close to the stellar radial direction (but
having intensities in the gauss range), in contradiction with the
observational results \cite{3, 4}  
obtained from spectral lines whose lower
level is intrinsically unpolarizable. 

\section{Multilevel modelling of the Hanle and Zeeman effects: 
diagnostics of chromospheric magnetic fields}

The physical interpretation of weak polarization signals 
requires to calculate the polarization of the atomic or molecular levels
within the framework of a rigorous theory for
the generation and transfer of polarized radiation. A suitable
theory for many spectral lines of diagnostic interest is the
density matrix polarization transfer theory of Landi Degl'Innocenti,
which is based on the Markovian assumption of complete frequency 
redistribution \cite{14, 15}.
This theory provides a physically consistent description of scattering phenomena
if the spectrum of the pumping radiation is flat across a sufficiently large
frequency range $\Delta\nu$ \cite{16}\footnote{The required extension of this 
$\Delta\nu$-interval depends on whether or not
coherences among Zeeman sublevels of {\it different} $J-$levels
can be neglected \cite{16}. 
If they need to be taken into account 
(as it occurs, {\em e.g.}, with the He {\sc i} ${\rm D}_3$ multiplet at 5876 \AA $\,$) 
then $\Delta\nu$ has to be of the order of the frequency range
of the multiplet. However, if such coherences
can be neglected (as it happens, {\em e.g.}, when modelling the Hanle effect in the
Ca {\sc ii} IR triplet) then $\Delta\nu$ needs to be only larger than
the inverse lifetime of the atomic levels.}.

The theoretical modelling of the He {\sc i} 10830 \AA $\,$ multiplet
in solar prominences and filaments (see 
the solid line of Fig. 1)
is based on the density matrix theory \cite{14, 15, 39}.
Trujillo Bueno {\em et al.} \cite{39}, have assumed a slab of He {\sc i} atoms 
lying at about 40'' above the
solar photosphere, from where it is illuminated by unpolarized 
and spectrally-flat radiation.
They have adopted a realistic multiterm model atom for He {\sc i} 
described in the incomplete Paschen-Back effect regime.
They also take into account coherences among magnetic sublevels
of each $J$-level, and between magnetic sublevels of the different
$J$-levels of each term (because they are important for some
terms of the model atom like {\em e.g.}, 
for the upper term of the ${\rm D}_3$ multiplet).
From the fitting to the spectropolarimetric observation
of the disk-center filament (open circles of Fig. 1)
they infer a magnetic field of 20 gauss and inclined
by about 105 degrees with respect to the radial direction through the observed
point. The agreement with the spectropolarimetric observation
is remarkable. It demonstrates that a very significant amount of the
atomic polarization that is induced by optical 
pumping processes in the metastable $2^3S_1$ lower-level 
survives the partial Hanle-effect
destruction due to horizontal magnetic fields with intensities
in the gauss range, and produces sizable linear polarization 
signals. 

As is well known, prominences
and filaments are located tens of thousands of kilometers above the
solar photosphere and their confining magnetic field does {\it not}
have a random azimuthal component within the spatio-temporal resolution element
of the observation. Therefore, one may ask whether the above-mentioned
belief can be safely applied to the solar chromosphere, where the
degree of anisotropy of the pumping radiation is significantly lower and the magnetic
fields may have a more complex topology.
This issue is investigated for the Ca {\sc ii} IR triplet 
by the authors in \cite{20, 38}.

Firstly, we have considered the zero magnetic field reference case and demonstrated that
the ``enigmatic'' relative $Q/I$ amplitudes 
(among the three lines) observed by Stenflo {\em et al.} \cite{31}, 
are the natural consequence of the existence of a sizable amount of atomic
polarization in the metastable levels $^2{\rm D}_{3/2}$ and $^2{\rm D}_{5/2}$
(which are the lower-levels of the Ca {\sc ii} IR triplet).
Secondly, we have investigated the Hanle effect in the IR triplet
at 8498, 8542 and 8662 \AA $\,$ considering a realistic
multilevel atomic model. 
Figure 4 is one of our most recent and interesting results,
which we will describe in full detail in forthcoming publications. 
It shows the fractional linear
polarization calculated at $\mu=0.1$ (about 5'' from the limb)
assuming magnetic fields of given inclination, but with a random
azimuthal component within the spatio-temporal resolution element of the observation.

\viewin \def\figfile{jtbueno_4}
 \begin{figure}[t]
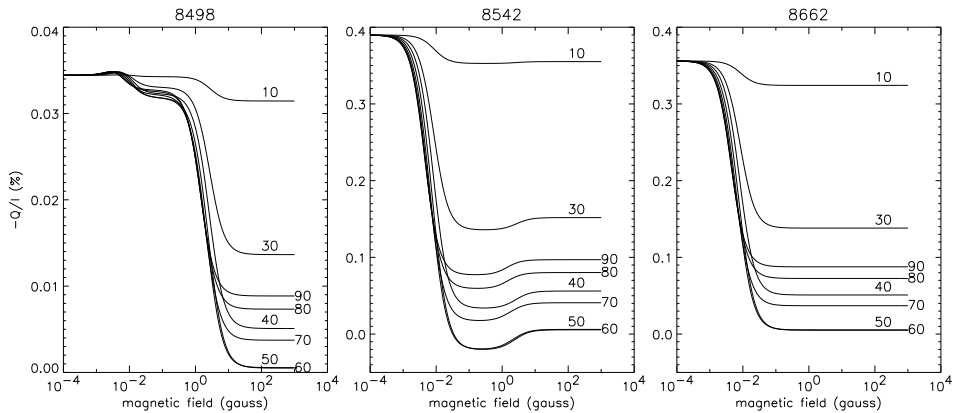

 \plotfig{130mm}{130mm}
   \caption[]{\label{fig:btodd} \captiontext{
The fractional linear polarization of the Ca {\sc ii} 
IR triplet calculated
at $\mu=0.1$ (about 5'' from the limb) in an isothermal atmosphere
with T=6000 K. Each curve corresponds to the indicated
inclination ($\theta_B$) of the assumed random-azimuth magnetic field.}}
\end{figure} \viewout

The results of this figure indicate that, basically, there are two
magnetic-field topologies (assuming that the
magnetic field lines have a random azimuthal
component over the spatio-temporal resolution element
of the observations) for which the limb polarization
signals of the 8542 and 8662 \AA $\,$ lines can have
amplitudes with $Q/I\,\ga\,0.1\%$ ({\em i.e.}, of the order of the observed ones).
As one could have expected, the first topology 
corresponds to magnetic fields with inclinations
$\theta_B\,\la\,30^\circ$. The second corresponds 
to magnetic fields which are practically
parallel to the solar surface, {\em i.e.}, ``horizontal'' fields 
with $80^{\circ}\,\la\,{\theta_B}\,\la\,100^{\circ}$. This demonstrates that
a significant amount of the atomic polarization that is induced by optical 
pumping processes in the metastable $^2{\rm D}_{3/2}$ lower-level 
survives the partial Hanle-effect
destruction produced by non-resolved canopy-like
horizontal fields with intensities
in the gauss range, and generates significant linear polarization 
signals via the dichroism mechanism.

The spectropolarimetric observation of Fig. 2 is only one example among many
other different cases of our GCT observations.
The sizable Stokes $V/I$ signal of Fig. 2 indicates that we were observing here
a moderately magnetized region close to the solar limb. 
Within the framework of the CRD theory of line formation
(see \cite{15}), this particular
observation of Fig. 2 cannot be modelled assuming a random azimuth magnetic field,
otherwise Stokes $U$ would have been undetectable. It would be of interest to confirm
with other telescopes the detection of that significant $U/I$ signal for the 8662 \AA$\,$ 
Ca {\sc ii} line, because it can only be due to the existence
of quantum interferences (coherences!) among the Zeeman sublevels of the
metastable $^2{\rm D}_{3/2}$ lower-level (see Section 7 in \cite{34}).
For this particular observation of Fig. 2
a good fit can be obtained assuming deterministic magnetic fields
with intensities in the gauss range and having inclinations $\theta_B\,{\la}\,30^{\circ}$
(see the multilevel Hanle and Zeeman modelling of Fig. 5). 
In any case, observations of more ``quiet'' and more ``active'' solar limb 
regions have been performed.
In some regions $Q$ is detected, whereas $U{\approx}0$ and/or $V{\approx}0$. 
In other
regions $V$ is detected, but $Q{\approx}U{\approx}0$. 
The physical interpretation of
these spectropolarimetric observations 
in terms of the Hanle and Zeeman effects is giving us
valuable clues about the intensities 
and magnetic field topologies in different regions close
to the solar limb.

\begin{figure}
  \centerline{\epsfig{figure=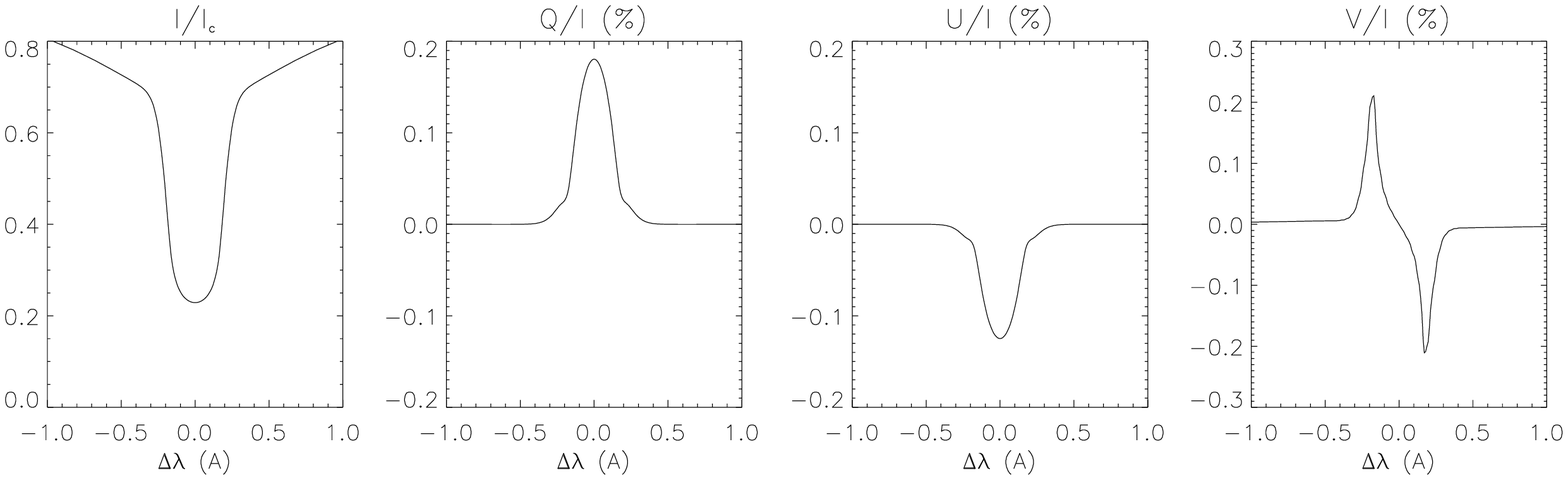, width=14.25cm, height=6.0cm, angle=0}}
  \caption{\label{fig:five} 
	The emergent Stokes parameters of the Ca {\sc ii} 8662 \AA$\,$ line
calculated at $\mu=0.1$ in the FAL-C semi-empirical model. We have assumed
a deterministic magnetic field of 20 gauss that is inclined by $25^{\circ}$ 
with respect to the radial direction. This figure is to be compared with 
the observational results of Fig. 2.}
\end{figure}  

\section{Concluding remarks}

The physical origin of the ``enigmatic'' linear polarization signals
observed in a variety of chromospheric lines is the existence
of atomic polarization in their metastable lower-levels,
which permits the operation of a {\it dichroism} mechanism that has nothing to do
with the transverse Zeeman effect. Therefore, the absorption process itself
plays a key role in producing the linear polarization signals
observed in the ``quiet'' solar chromosphere as well as in solar filaments.

The population imbalances and coherences among the Zeeman sublevels
of such {\it long-lived} atomic levels
can be sufficiently significant in the presence of horizontal magnetic
fields having intensities in the gauss range
(see, however, \cite{40} concerning the 
very particular case of the `enigmatic' sodium D$_1$ line).
Therefore, in general, one should not
feel obliged to conclude that the magnetic fields
throughout the ``quiet'' solar
chromosphere have to be either extremely low ({\em i.e.}, 
with intensities $B{\la}10$ mG),
or, alternatively, oriented preferentially along the radial
direction. 
The physical interpretation of our spectropolarimetric observations
of chromospheric lines in terms of the Hanle and Zeeman effects
indicates that the magnetic field topology 
may be considerably more complex, having both moderately inclined and 
practically horizontal field lines with 
intensities above the milligauss range. 
A physically plausible scenario 
that might lead to polarization signals in agreement
with the observations is that
resulting from the superposition of miriads of different loops of magnetic field lines
connecting opposite polarities. This suggested magnetic field
topology is somehow reminiscent of the magnetic structure model of the
``quiet'' transition region proposed by Dowdy {\em et al.} \cite{6}, 
but scaled down to the spatial dimensions of the solar chromosphere. 

\acknowledgments
We are grateful to the following scientists
for their collaboration and for useful discussions: 
Manolo Collados, Olaf Dittmann, Egidio Landi
Degl'Innocenti, Valentin Mart\'\i nez Pillet, 
Laura Merenda, Frederic Paletou and Meir Semel. 
This work is part of the EC-TMR European Solar Magnetometry
Network and has been partly funded by the Spanish 
Ministerio de Ciencia y Tecnolog\'\i a through project AYA2001-1649.

\end{document}